\documentstyle[prl,aps,epsf]{revtex}
\input epsf
\begin{document}
\draft
\twocolumn[
\hsize\textwidth\columnwidth\hsize\csname @twocolumnfalse\endcsname

\title{Impurity spin magnetization of thin Fe doped Au films}

\author{E.~Seynaeve, K.~Temst, F.~G.~Aliev$^{\dagger}$, C.~Van~Haesendonck}
\address{Laboratorium voor Vaste-Stoffysica en Magnetisme,
Katholieke Universiteit Leuven, \\ Celestijnenlaan 200 D, B-3001 Leuven,
Belgium}
\author{V.~N.~Gladilin$^*$, V.~M.~Fomin$^{*,**}$,
J.~T.~Devreese$^{**,\sharp}$}
\address{Theoretische Fysica van de Vaste Stof, Departement Natuurkunde,
Universiteit Antwerpen (UIA), \\ Universiteitsplein 1, B-2610 Antwerpen,
Belgium}

\date{\today}
\maketitle

\begin{abstract}
In order to probe the influence of
the surface-induced anisotropy on the impurity spin magnetization,
we measure the anomalous Hall effect in thin AuFe
films at magnetic fields up to $15 \, {\rm T}$.
The observed suppression of the anomalous Hall resistivity
at low fields as well as the appearance of a
minimum in the differential Hall resistivity
at higher fields can be explained by our theoretical model which takes
into account the influence of a polycrystalline film structure on
the surface-induced anisotropy.
Our results imply that the apparent
discrepancy between different experimental results
for the size effects in dilute magnetic alloys can be linked to a
different microstructure of the samples.

\end{abstract}

]

In very dilute magnetic alloys finite size effects may occur for
sample dimensions comparable to the size of the Kondo screening
cloud \cite{Affleck}. While some experiments
\cite{Giordano} revealed a considerable decrease of the logarithmic Kondo
anomaly in the resistivity of thin films and narrow wires already at
the $\mu {\rm m}$ scale, other experiments \cite{Chandrasekhar} showed
an almost constant Kondo anomaly for wire widths down to $40 \, {\rm
nm}$. Recent theoretical calculations indicated the complex, dynamical
nature of the screening cloud, implying that the simple picture of a
static, spherically symmetric screening cloud is not correct
\cite{Affleck,Bergmann}.

Zawadowski {\em et al.} \cite{Zawadowski} linked the size dependent
Kondo scattering to a surface-induced anisotropy of
the magnetic impurity spins. Due to the interaction of
the impurities with the conduction
electrons, which suffer from the spin-orbit scattering by the non-magnetic
host atoms, the impurity spins tend to be aligned parallel to the
sample boundaries \cite{Fomin}. The surface-induced anisotropy should
remain active for more concentrated spin glass alloys. Finite size
effects have indeed been observed in the resistivity of spin
glasses with reduced dimensions \cite{Parpia,Neut}.

Finite size effects should also be observable in the impurity spin
magnetization. Earlier magnetization experiments on single film
\cite{Vloeberghs} and on multilayered \cite{Kenning} spin glasses had
linked a depression of the freezing temperature $T_{f}$ to a lower critical
dimensionality for the spin glass transition. Here, we show that measuring
the magnetic field dependence
of the anomalous Hall effect in thin films of a AuFe spin glass
\cite{Vloeberghs} provides a powerful method to probe the influence of
the surface-induced anisotropy on the Fe spin magnetization. At low
magnetic fields, the magnetization signal related to the spin glass
freezing is strongly suppressed when compared to the bulk
behavior.  The appearance of an extra magnetization
signal at higher fields can be linked to a re-orientation of impurity spins
which are
blocked by the surface-induced anisotropy at lower fields. The low field
magnetization is further suppressed when moving
the Fe doping towards the film surface. We have developed a theoretical
model which takes into account the polycrystalline film
structure. The presence of additional
internal surfaces at the grain boundaries introduces a particular
spatial dependence of the spin anisotropy which is essential for
understanding the Hall effect data.  Our model provides an explanation
for the apparent discrepancies between the different experiments which
have probed the size dependence of the Kondo and spin glass
resistivity.

Thin films of AuFe alloys have been prepared by codeposition on
oxidized silicon wafers of Au ($99.9999\%$ purity) and Fe ($99.99\%$
purity) in an ultra high vacuum molecular beam epitaxy growth chamber.
A multiterminal sample geometry is obtained by depositing the films through
a contact mask. Apart from $30 \, {\rm nm}$ thick Fe doped Au
films and pure Au reference films, we have also prepared Au/AuFe/Au and
AuFe/Au/AuFe trilayers.  For the $30 \, {\rm nm}$ thick trilayer
samples the central layer has a thickness of $15 \, {\rm nm}$, while
the two outer layers have a thickness of $7.5 \, {\rm nm}$.  The Hall
resistance as well as the longitudinal resistance are measured with an
ac resistance bridge in perpendicular magnetic fields up to $15 \, {\rm
T}$.

\begin{figure}
\protect\centerline{\epsfbox{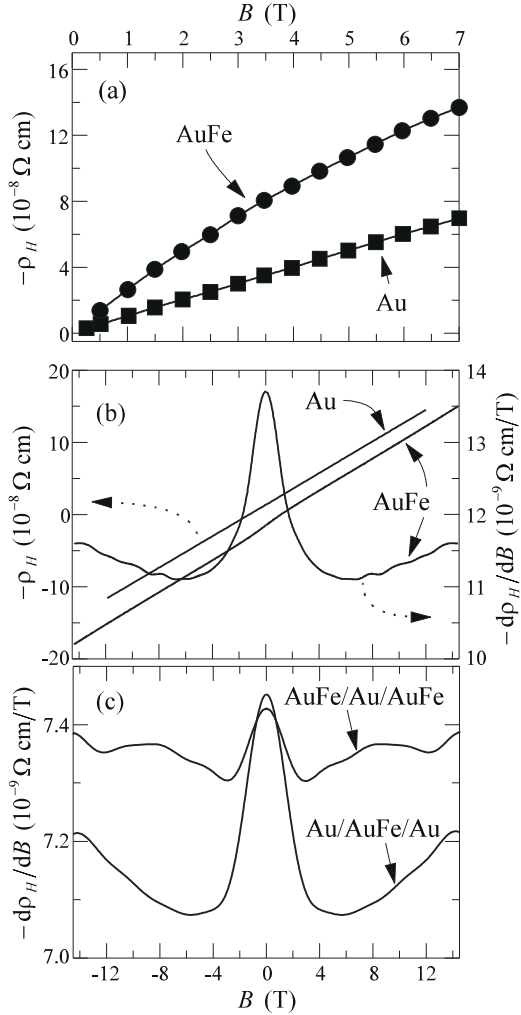}}
\caption{(a) Magnetic field dependence of the Hall resistivity
for a bulk polycrystalline Au sample and a bulk $0.98 \, {\rm at.\%}$
AuFe sample at $T = 4.2 \, {\rm K}$ (from Ref. \protect\cite{McAlister}).
(b) Magnetic field
dependence of the Hall resistivity for a pure Au film and for a
$2 \, {\rm at.\%}$
AuFe film at $T = 1 \, {\rm K}$. Both films have a thickness of
$30 \, {\rm nm}$. For the AuFe film the differential Hall resistivity
$- {\rm d} \rho_{H} / {\rm d} B$ is also shown. (c) Magnetic field
dependence of the differential Hall resistivity for the Au/AuFe/Au and
AuFe/Au/AuFe
trilayers (Fe concentration is $3.5 \, {\rm at.\%}$ for the AuFe) at $T =
1 \, {\rm K}$.}
\label{Fig.1}
\end{figure}

Adding Fe impurity spins to pure Au causes the appearance of a
non-linear anomalous Hall component. Figure~\ref{Fig.1}(a) shows
typical Hall effect data at $T = 4.2 \, {\rm K}$ which have been
obtained by McAlister and Hurd \cite{McAlister} for polycrystalline pure
bulk Au as well as for a $0.98 \, {\rm at.\%}$ AuFe bulk spin glass. The
anomalous Hall resistivity varies proportional to
the Fe spin magnetization and its temperature dependence at low fields
reveals a sharp peak near the freezing temperature $T_{f}$. In
agreement with direct magnetization measurements, the peak is washed
out at higher magnetic fields \cite{McAlister,Ul-Haq}.

As illustrated in Fig.~\ref{Fig.1}(b), the anomalous Hall effect is
much weaker in a thin $2 \, {\rm at.\%}$ AuFe spin glass film when
compared to the bulk material. Although the Fe con- \linebreak
\noindent centration is about
twice as large as in Fig.~\ref{Fig.1}(a), deviations from the classical
linear Hall resistivity of the pure Au film are very small. In order to
make the non-linear anomalous Hall component more clearly visible, we
also plot in Fig.~\ref{Fig.1}(b) the differential Hall resistivity $-
{\rm d} \rho_{H} / {\rm d} B$.  The initial variation of the
differential Hall resistivity at low fields reflects the destruction of
the spin glass state due to the alignment of the impurity spins. The
magnetic response resulting from this alignment has been reduced by an
order of magnitude when compared to the bulk material. As discussed in
detail below, we can attribute the reduced amplitude of $- {\rm
d}\rho_{H} / {\rm d} B$ at low fields as well as the appearance of a
minimum at higher fields to the surface-induced anisotropy.

The strength of the anisotropy of an impurity spin is predicted to increase
when the impurity approaches the sample surface \cite{Zawadowski}.
Consequently, the anisotropy effects should be less pronounced for
a Au/AuFe/Au trilayer when compared to a AuFe/Au/AuFe
trilayer. This is confirmed in Fig.~\ref{Fig.1}(c) for AuFe layers with
an Fe concentration of $3.5 \, {\rm at.\%}$. Although both the total
thickness and the Fe content are the same, the non-linear behavior at
low fields is clearly enhanced when moving the Fe impurities away from
the surface. The reduced amplitude of $- {\rm d} \rho_{H} / {\rm d} B$
for the trilayers when compared to a single AuFe film (see
Fig.~\ref{Fig.1}(b)) results from a shortcircuiting by the pure Au layer
with a much lower resistivity.

In order to calculate the non-linear impurity magnetization in
polycrystalline AuFe films, we extended the theory of the
surface-induced anisotropy \cite{Zawadowski,Fomin} to the case of thin
films consisting of small grains.  Following Ref.~\onlinecite{Fomin},
the Hamiltonian ${\cal H}_{\rm an}$, which describes the surface-induced
anisotropy for an impurity spin in an
isolated brick-shaped grain, can be written as
\begin{eqnarray}
&&{\cal H}_{\rm an}={\cal A}\sum_{\alpha,\beta}{\cal B}_{\alpha \beta}
S_{\alpha}S_{\beta} \quad  (\alpha,\beta=x,y,z).
\label{han}
\end{eqnarray}
The $S_{\alpha}$ are the operators for the
components of the impurity spin ${\bf S}$.
In contrast to a semi-infinite sample \cite{Zawadowski},
the strength of the anisotropy has to be described in terms of a matrix
with elements ${\cal A}{\cal B}_{\alpha\beta}$.
Moreover, the calculated matrix elements ${\cal
B}_{\alpha \beta}$ are complicated functions of the dimensions
($a_{x}$, $a_{y}$, $a_{z}$) of the grain and of the impurity position.
The material dependent constant ${\cal A}$ should range between 0.01
and $1 \, {\rm eV}$ for dilute AuFe alloys \cite{Zawadowski,Fomin}.
Taking the $z$-axis parallel to the magnetic field, the magnetization of an
Fe spin is given by
\begin{eqnarray}
\langle S_z \rangle=-\frac{1}{2\mu_{\rm B}{\cal Z}}
\sum_{k=1}^{5}{\rm exp}\left(-\frac{{\cal E}_k}{k_{\rm B}T}\right)
\frac{{\rm d}{\cal E}_k}{{\rm d} B}, \label{avS2}
\end{eqnarray}
where $\mu_{\rm B}$ is the Bohr magneton, and
${\cal Z}=\sum_{k=1}^{5}{\rm exp}(-{\cal E}_k/k_{\rm B}T$).
The index $k = 1,...,5$ labels the
roots ${\cal E}_k$ of the secular equation
\begin{eqnarray}
&&\left|{\cal H}_{S_z^{\prime}S_z^{}}-{\cal
E}\delta_{S_z^{\prime}S_z^{}} \right|=0\qquad
(S_z^{},S_z^{\prime}=-2,-1,0,1,2) \label{sec}
\end{eqnarray}
with the Hamiltonian
${\cal H}=-2\mu_{\rm B} S_zB + {\cal H}_{\rm an}$.

\begin{figure}
\protect\centerline{\epsfbox{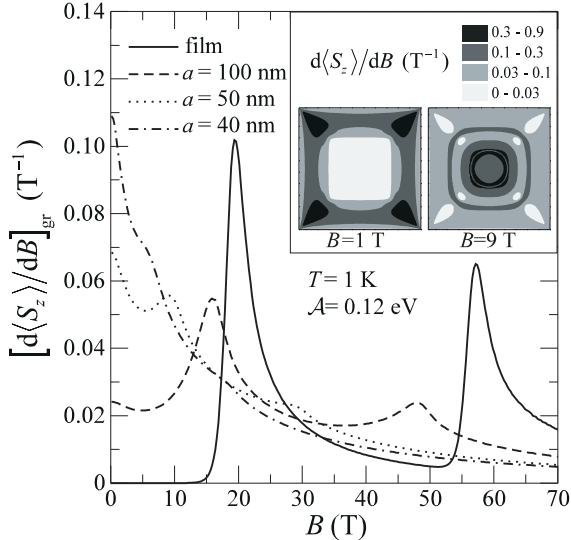}}
\caption{Calculated magnetic field dependence of the differential
magnetization $\left[{\rm d}\langle S_{z}\rangle/{\rm d}B\right]_{\rm gr}$
for grains with height $a_{z} = 30 \, {\rm nm}$ and lateral dimensions
$a_{x} = a_{y} = a$. Inset: Spatial distribution of the differential
magnetization for a cross-section perpendicular to the magnetic field at
the center of a grain with $a = 50 \, {\rm nm}$.}
\label{Fig.2}
\end{figure}

In Fig.~\ref{Fig.2} the calculated differential magnetization
$\left[{\rm d}\langle S_{z}\rangle/{\rm d}B\right]_{\rm gr}$ (the square
brackets with subscript `gr'
correspond to an average over impurity positions within a grain) is
shown as a function of the magnetic field for different lateral dimensions
of the grains.  The height
$a_{z}$ of the grains is fixed at $30 \, {\rm nm}$, i.e., the thickness
of the AuFe film shown in Fig.~\ref{Fig.1}(b). In the limiting case of
a {\it single-crystal film} ($a_{x}, a_{y} \rightarrow \infty$), the
impurity spin states at $B=0$ are known \cite{Zawadowski} to be the
eigenstates of $S_{z}$, the energy eigenvalues being proportional to
$S_{z}^{2}$. At low temperatures, only the state with $S_{z} = 0$ will
be populated.  Hence, the impurity spin does not respond to a weak
magnetic field.  With increasing $B$ the energy level with $S_{z} = 1$
becomes lower than that with $S_{z} = 0$. This gives rise to the first
peak in the field dependence of
$\left[{\rm d}\langle S_{z}\rangle/{\rm d}B\right]_{\rm gr}$.
The second peak appears when the state with $S_{z} =
2$ becomes the ground state. The theoretical result for the
single-crystal film in Fig.~\ref{Fig.2} clearly fails to describe our
experimental results, since the calculated impurity spin magnetization
remains zero at low magnetic fields (all spins are blocked parallel to
the surface).

In Ref.~\onlinecite{Fomin} it was already shown that in narrow wires
the presence of differently oriented surfaces leads to a rather
intricate behavior of the impurity-spin anisotropy when compared to the
case of a film. In {\it brick-shaped grains}, the competing influence of
mutually perpendicular surfaces on the magnetic anisotropy gives rise to a
partial cancellation of the anisotropy effect, resulting in a shift of
the peaks in $\left[{\rm d}\langle S_{z}\rangle/{\rm d}B\right]_{\rm gr}$
towards lower fields. Moreover, there exist specific locations within
the grains where the magnetic anisotropy energy becomes negligibly
small. At these locations the bulk behavior of the impurity spins is
restored. Hence, a limited number of spins contribute to the
spin glass freezing and are sensitive to very weak magnetic fields. In the
inset of Fig.~\ref{Fig.2} we illustrate the large difference in the
predicted response of spins located near the corners of the grains when
compared to spins located more towards
the center of the grains.
For grains with lateral sizes larger than (but still comparable to)
the height, our theoretical model predicts that
a minimum in $\left[{\rm d}\langle S_{z}\rangle/{\rm d}B\right]_{\rm gr}$
can appear at relatively small magnetic fields. For those
grains the initial part of the theoretical curves is
similar to the measured differential Hall resistivity
shown in Fig.~\ref{Fig.1}(b) and in Fig.~\ref{Fig.1}(c).
At higher temperatures ($T > 10 \, {\rm K}$) the anomalous Hall effect
signal gradually weakens.
A consistent description of our anomalous Hall data requires to assume
a large value ${\cal A} = 0.12 \, {\rm eV}$ in Eq.~(1). This implies that
a larger fraction of the alignment of the
Fe spins by the magnetic field occurs at extremely
high fields which are not accessible in the experiment.

\begin{figure}
\protect\centerline{
\epsfbox{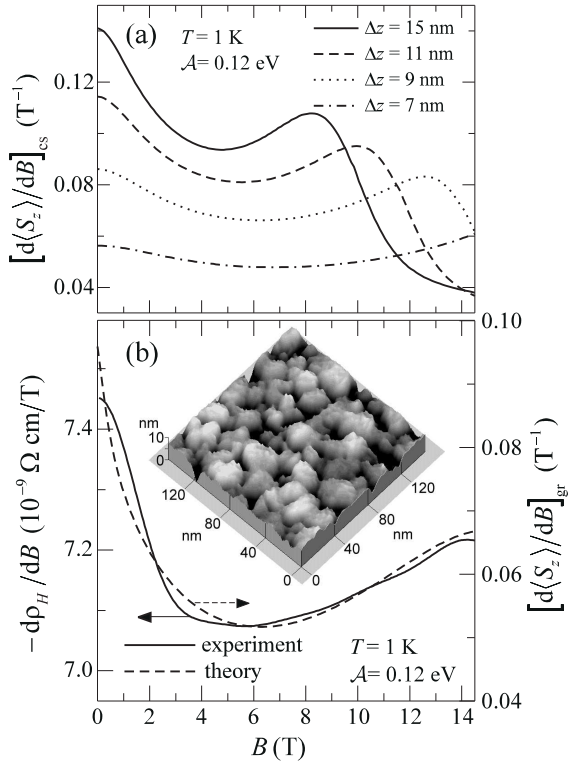}}
\caption{(a) Calculated magnetic field dependence of the
differential magnetization
$\left[{\rm d}\langle S_{z}\rangle/{\rm d}B\right]_{\rm cs}$ for a grain
after averaging over a cross-section perpendicular to the magnetic field.
Different curves are for different distances $\Delta z$ to the top or bottom
of the grain with a height of $30 \, {\rm nm}$ and lateral size of
$50 \, {\rm nm}$.
(b) Comparison between the differential Hall resistivity
$-{\rm d} \rho_{H} / {\rm d} B$ measured for
the Au/AuFe/Au trilayer (solid curve) and the calculated
differential magnetization
$\left[{\rm d}\langle S_{z}\rangle/{\rm d}B\right]_{\rm gr}$
for an ensemble of Au/AuFe/Au grains with
a height of $30 \, {\rm nm}$ and
a thickness of the central AuFe layer of $15 \, {\rm nm}$ (dashed curve).
For the lateral grain size $a$ a lognormal distribution is assumed (see text).
The inset shows an STM image of the surface of the Au/AuFe/Au trilayer.}
\label{Fig.3}
\end{figure}

Figure~\ref{Fig.3}(a) shows the calculated field
dependence of an average
$\left[{\rm d}\langle S_{z}\rangle/{\rm d}B\right]_{\rm cs}$
over the positions of the impurities located within a cross-section of a
grain perpendicular to the magnetic field.
The different curves correspond to different distances
$\Delta z$ from the cross section to the
top (or bottom) of the grain. The theoretical curves can be
compared to the experimental field dependence of $-{\rm d} \rho_{H} /
{\rm d} B$ for the Au/AuFe/Au and the AuFe/Au/AuFe trilayer shown in
Fig.~\ref{Fig.1}(c). The low field variation is much weaker for the
AuFe/Au/AuFe trilayer. This is in agreement with our theoretical
result that the surface-induced anisotropy becomes stronger for Fe
spins which are closer to the top (or bottom) of a grain.

The inset of Fig.~\ref{Fig.3}(b) shows a scanning tunneling microscopy
(STM) image of the surface of the Au/AuFe/Au sample, revealing a rather
broad distribution of grain sizes. In order to quantitatively fit the
experimental data we therefore calculate the magnetization for an ensemble
of grains of different lateral size $a$,
distributed lognormally with a statistical median $\bar a$ and a standard
deviation $\sigma$.
In Fig.~\ref{Fig.3}(b),
the measured differential Hall
resistivity $-{\rm d} \rho_{H} / {\rm d} B$ for the Au/AuFe/Au trilayer
is compared to the field dependence of
$\left[{\rm d}\langle S_{z}\rangle/{\rm d}B\right]_{\rm gr}$,
calculated for an ensemble of
Au/AuFe/Au grains with
$\bar a=80\, {\rm nm}$ and $\sigma = 1.5$.
We note that the value of $80 \, {\rm nm}$ is larger
than the typical grain size inferred from the STM images ($20 \, {\rm nm}$).
Our theoretical approach with isolated grains is strictly valid only when
elastic defect scattering at the grain boundaries fully destroys the
anisotropy which is caused by the
conduction-electron-mediated interaction of an Fe spin with host atoms
from neighboring grains. Using a larger grain size for the calculations
allows to take into account that a fraction of the electrons still moves
ballistically between adjacent grains.

We conclude that our theoretical model for the surface-induced
anisotropy of magnetic impurities in small metallic grains allows a
quantitative description of the anomalous Hall resistivity in
polycrystalline AuFe spin glass films.  The small impurity spin
magnetization at low magnetic field also provides an explanation for
the small amplitude of the spin glass resistivity in thin AuFe films
\cite{Neut}. It is clear that the surface-induced anisotropy strongly
affects the spin glass freezing, implying that the analysis
of previous
measurements, which investigated the reduction of the freezing
temperature $T_{f}$ in thin films \cite{Vloeberghs} and multilayers
\cite{Kenning}, should be revised. At this point it is not clear in how
far this reduction, which also occurs for the thin film samples
discussed in this Letter, is caused by an intrinsic finite size effect
or can be totally accounted for by the surface-induced anisotropy.

For Kondo alloys with a small impurity content
\linebreak ($\sim 100 \, {\rm ppm}$) we are no longer able to measure the
anomalous Hall effect. We expect the
surface-induced anisotropy to still produce a strong reduction of the
impurity spin magnetization, resulting in a
suppression of the Kondo resistivity. Since the surface-induced blocking of
a magnetic impurity spin is sensitive to the specific polycrystalline sample
structure, the apparent
discrepancy between different experimental results
\cite{Giordano,Chandrasekhar} for the size dependence of the Kondo
resistivity can be linked to a different microstructure of the samples.

The collaboration between the universities of Antwerpen and Leuven has been
supported by the Fund for Scientific Research -- Flanders (Belgium) as well as
by the Belgian Inter-University Attraction Poles research program
on Reduced Dimensionality Systems (IUAP No.~4/10). Additional support has
been obtained in Leuven from the Flemish Concerted Action (GOA) research program
and in Antwerpen from the Scientific Fund (BOF) of the Universiteit Antwerpen.

%

%
%

\end{document}